\documentclass[longauth, traditabstract]{aa} 
\usepackage{graphicx}
\usepackage{txfonts}
\usepackage{natbib}
\def\aj{AJ}%
\def\apj{ApJ}%
\def\apjl{ApJ}%
\def\apjs{ApJS}%
\def\aap{A\&A}%
\def\new#1{{#1}}

\begin{document}

\newcommand{\htwopo}{\mbox{\emph{ortho}-H$_2$O$^+$}}
\newcommand{\htwopp}{\mbox{\emph{para}-H$_2$O$^+$}}
\newcommand{\htwop}{\mbox{H$_2$O$^+$}}

\title{\textit{Herschel} observations of ortho- and para-oxidaniumyl (H$_2$O$^+$) in spiral arm clouds toward Sgr~B2(M) \thanks{\textit{Herschel} is an ESA space observatory
    with science instruments provided by European-led Principal
    Investigator consortia and with important participation from
    NASA.}}


\author{
P.~Schilke,\inst{1,2}
C.~Comito,\inst{2}
H.~S.~P. M\"uller,\inst{1}
E.~A.~Bergin,\inst{3}
E.~Herbst,\inst{14}
D.~C.~Lis,\inst{4}
D.~A.~Neufeld,\inst{20}
T.~G.~Phillips,\inst{4}
T.~A.~Bell,\inst{4}
G.A.~Blake,\inst{5}
S.~Cabrit,\inst{24}
E.~Caux,\inst{6,7}
C.~Ceccarelli,\inst{8}
J.~Cernicharo,\inst{9}
N.~R.~Crockett,\inst{3}
F.~Daniel,\inst{9,10}
M.-L.~Dubernet,\inst{11,12}
M.~Emprechtinger,\inst{4}
P.~Encrenaz,\inst{10}
M.~Gerin,\inst{10}
T.~F.~Giesen,\inst{1}
J.~R.~Goicoechea,\inst{9}
P.~F.~Goldsmith,\inst{13}
H.~Gupta, \inst{13}
C.~Joblin,\inst{6,7}
D.~Johnstone,\inst{15}
W.~D.  Langer\inst{13} 
W.~B. Latter\inst{16}
S.~D.~Lord,\inst{16}
S.~Maret,\inst{8}
P.~G.~Martin,\inst{17}
G.~J.~Melnick,\inst{18}
K.~M.~Menten,\inst{2}
P.~Morris,\inst{16}
J.~A.~Murphy,\inst{19} 
V.~Ossenkopf,\inst{12,21}
L.~Pagani,\inst{24}
J.~C.~Pearson,\inst{13}
M.~P\'erault,\inst{10}
R.~Plume,\inst{22}
S.-L.~Qin,\inst{13}
M.~Salez, \inst{24}
S.~Schlemmer,\inst{1}
J.~Stutzki,\inst{1}
N.~Trappe,\inst{19}
F.~F.~S.~van der Tak,\inst{21}
C.~Vastel,\inst{6,7}
S.~Wang,\inst{3}
H.~W.~Yorke,\inst{13}
S.~Yu,\inst{13}
N. Erickson,\inst{25}
F.W. Maiwald,\inst{13}
J. Kooi,\inst{4}
A. Karpov,\inst{4}
J.~Zmuidzinas,\inst{4}
A. Boogert, \inst{4}
R. Schieder, \inst{1}
\and
P. Zaal\inst{21}
}
\institute{I. Physikalisches Institut, Universit\"at zu K\"oln,
              Z\"ulpicher Str. 77, 50937 K\"oln, Germany\\
\email{schilke@ph1.uni-koeln.de}
\and Max-Planck-Institut f\"ur Radioastronomie, Auf dem H\"ugel 69, 53121 Bonn, Germany 
\and Department of Astronomy, University of Michigan, 500 Church Street, Ann Arbor, MI 48109, USA 
\and California Institute of Technology, Cahill Center for Astronomy and Astrophysics 301-17, Pasadena, CA 91125 USA
\and  California Institute of Technology, Division of Geological and Planetary Sciences, MS 150-21, Pasadena, CA 91125, USA
\and Centre d'\'etude Spatiale des Rayonnements, Universit\'e de Toulouse [UPS], 31062 Toulouse Cedex 9, France
\and CNRS/INSU, UMR 5187, 9 avenue du Colonel Roche, 31028 Toulouse Cedex 4, France
\and Laboratoire d'Astrophysique de l'Observatoire de Grenoble, 
BP 53, 38041 Grenoble, Cedex 9, France.
\and Centro de Astrobiolog\'ia (CSIC/INTA), Laboratiorio de Astrof\'isica Molecular, Ctra. de Torrej\'on a Ajalvir, km 4
28850, Torrej\'on de Ardoz, Madrid, Spain
\and LERMA, CNRS UMR8112, Observatoire de Paris and \'Ecole Normale Sup\'erieure, 24 Rue Lhomond, 75231 Paris Cedex 05, France
\and LPMAA, UMR7092, Universit\'e Pierre et Marie Curie,  Paris, France
\and  LUTH, UMR8102, Observatoire de Paris, Meudon, France
\and Jet Propulsion Laboratory,  Caltech, Pasadena, CA 91109, USA
\and Departments of Physics, Astronomy and Chemistry, Ohio State University, Columbus, OH 43210, USA
\and National Research Council Canada, Herzberg Institute of Astrophysics, 5071 West Saanich Road, Victoria, BC V9E 2E7, Canada 
\and Infrared Processing and Analysis Center, California Institute of Technology, MS 100-22, Pasadena, CA 91125
\and Canadian Institute for Theoretical Astrophysics, University of Toronto, 60 St George St, Toronto, ON M5S 3H8, Canada
\and Harvard-Smithsonian Center for Astrophysics, 60 Garden Street, Cambridge MA 02138, USA
\and  National University of Ireland Maynooth. Ireland
\and  Department of Physics and Astronomy, Johns Hopkins University, 3400 North Charles Street, Baltimore, MD 21218, USA
\and SRON Netherlands Institute for Space Research, PO Box 800, 9700 AV, Groningen, The Netherlands
\and Department of Physics and Astronomy, University of Calgary, 2500
University Drive NW, Calgary, AB T2N 1N4, Canada
\and
University of Massachusetts, Astronomy Dept., 710 N. Pleasant St., LGRT-619E, Amherst, MA 01003-9305  U.S.A
\and
LERMA \& UMR8112 du CNRS, Observatoire de Paris, 61, Av. de l'Observatoire, 75014 Paris, France
}


\abstract{
\htwop{} has been observed in its \textit{ortho}- and \textit{para}- states toward the massive star forming core \object{Sgr~B2(M)}, located close to the Galactic center.  The observations show absorption in all spiral arm clouds between the Sun and Sgr~B2.  The average o/p ratio of \htwop{} in most velocity intervals is 4.8, which corresponds to a nuclear spin temperature of 21~K.  The relationship of this spin temperature to the formation temperature and current physical temperature of the gas hosting \htwop{} is discussed, but no firm conclusion is reached.  In the velocity interval 0-60 km s$^{-1}$, an \textit{ortho}/\textit{para} ratio of below unity is found, but if this is due to an artifact of contamination by other species or real is not clear. 

}

   \titlerunning{Oxidaniumyl toward Sgr~B2}
	\authorrunning{Schilke et al.}
   \maketitle
%

\section{Introduction}

Simple di- and triatomic molecules and ions are fundamental constituents of interstellar chemistry which eventually leads to the formation of complex molecules. Many of these species have ground state transitions at submillimeter- and THz wavelengths, and are therefore either difficult or not at all observable from the ground, yet they constitute the building blocks of chemistry, and are therefore fundamental to its understanding in various environments.   Among those are spiral arm clouds, located in the plane of the Galactic disk, where the line-of-sight toward a strong continuum source passes through by chance.  This setup allows sensitive absorption measurements, and the clouds have been observed by this method against the Sgr B2, W31c, W49, W51 and CasA millimeter continuum sources using  molecular species such as CO, HCN, HCO$^{+}$, CS, CN, SO, and c-C$_{3}$H$_{2}$ etc \citep[e.g.][]{Greaves1994, Tieftrunk1994, Greaves1996,  Menten2010, Gerin2010, Ossenkopf2010}. The results demonstrate that spiral arm clouds have low gas density and low excitation temperatures, and  represent diffuse and translucent clouds.

One of the best sources for these absorption studies is Sgr B2, located close to the Galactic center, $\sim$ 100 pc from Sgr A$^{\star}$ in projection, and one of the strongest submillimeter sources in the Galaxy \citep[e.g.][]{Pierce-Price2000}. The dense cores Sgr~B2(N) and Sgr~B2(M) within the cloud are at different evolutionary stages, and constitute well-studied massive star forming regions in our Galaxy. The flux ratio of the continuum between Sgr~B2(M) and Sgr~B2(N) is less than unity at 1 mm and rises at shorter wavelengths so that Sgr~B2(M) dominates above $\sim$ 500 GHz \citep{Goldsmith1990, Lis1991}.  Sgr~B2(M) also shows fewer molecular emission lines than Sgr~B2(N) \citep{Nummelin1998}, hence less confusion and therefore is better suited for absorption studies. The line-of-sight toward the Sgr~B2(M) continuum will pass \new{almost all the way to the center of our Galaxy}, providing a more complete census in studying physical and chemical conditions towards the Galactic center clouds and all spiral arm clouds simultaneously. HIFI, the Heterodyne Instrument for the Far-Infrared \citep{deGraauw2010} on board the {\it Herschel} Space Observatory \citep{Pilbratt2010} is an ideal instrument for making these observations.

\section{Observations}
Full spectral
scans of HIFI bands \new{1a, 1b, and 4b} towards Sgr~B2(M) ($\alpha_{J2000} = 17^h47^m20.35^s$ and $\delta_{J2000} =
-28^{\circ}23'03.0''$) have been carried out respectively on {\bf
  March 1, 2, and 5 2010}, providing coverage of the frequency range 479 through
637 GHz \new{and 1051 through 1121 GHz}.

HIFI Spectral Scans are carried out in Dual Beam Switch (DBS)
mode, where the DBS reference beams lie approximately 3$^{\prime}$
apart. The wide band spectrometer (WBS) is used as a back-end,
providing a spectral resolution of 1.1 MHz over a 4-GHz-wide
Intermediate Frequency (IF) band.  A HIFI Spectral Scan consists
of a number of double-sideband (DSB) spectra, tuned at different Local
Oscillator (LO) frequencies, where the spacing between one LO setting
and the next is determined by the ``redundancy'' chosen by the
observer \citet{Comito2002}. The molecular spectrum of Sgr~B2(M) in
\new{band 1a and 4b has been scanned
with a redundancy of 4, that of band 1b with a redundancy of 8,} which means that every
frequency has been observed respectively 4 and 8 times in each
sideband. Multiple observations of the same frequency at different LO
tunings are necessary to separate the lower-sideband (LSB) from the
upper-sideband (USB) spectrum.

The data have been calibrated through the standard pipeline released with
version 2.9 of HIPE \citet{Ott2010}, and subsequently exported to
CLASS\footnote{{\it Continuum and Line Analysis Single-dish Software},
 distributed with the GILDAS software, see http://www.iram.fr/IRAMFR/GILDAS.} using the HiClass task within HIPE. Deconvolution of
the DSB data into single-sideband (SSB) format has been performed on
CLASS. All the HIFI data presented here, spectral features \emph{and}
continuum emission, are deconvolved SSB spectra. Although both
horizontal (H) and vertical (V) polarizations have been obtained, we will show
only  H-polarization spectra. The intensity scale
is main-beam temperature, and results from applying a beam efficiency
correction of \new{0.69 for band 1a, 0.68 for band 1b, and 0.669 for
  band 4b} \citep{Roelfsema2010}.

\section{Spectroscopy of H$_2$O$^+$}\label{sec:spec}
Removal of an electron from oxidane, H$_2$O, also known 
as water, yields oxidaniumyl, H$_2$O$^+$. Its bond lengths and 
bond angle are slightly larger than \new{those of} H$_2$O, see e.g. 
\citet{H2O+_LMR_1986}. Quantum-chemical calculations 
\citep{H2O+_ai_1989} yielded a ground state dipole moment of 
$\sim$2.4~D, considerably larger than in H$_2$O. 
The transitions are of {\it b}-type, meaning 
$\Delta K_a \equiv \Delta K_ \equiv 1$~mod~2.
The electronic ground state changes from $^1A_1$ in the neutral 
to $^2B_1$ in the cation which leads to a reversal of the 
{\it ortho} and {\it para} levels with respect to water.
$K_a + K_c$ is even and odd for {\it ortho}- and 
{\it para}-H$_2$O$^+$, respectively.
The {\it para} levels do not show any hyperfine splitting 
while the {\it ortho} levels are split into three because 
of the $^1$H hyperfine structure. The strong lines have 
$\Delta F = \Delta J = \Delta N$, i.e. they do not involve a 
spin-flip. At low quantum numbers spin-flipping transitions 
have appreciable intensity.

Further details of the spectroscopy of H$_2$O$^+$ are discussed 
in the Appendix. Table~\ref{lab-data} provides calculated rest 
frequencies for the two rotational transitions discussed 
in the present investigation. Fig.~\ref{levels} show the lowest 
energy levels of H$_2$O$^+$ with allowed transitions.

\section{Results}
\begin{figure}
   \centering
   \includegraphics[bb=-50 100 600 600, width=0.75\columnwidth, angle=-90]{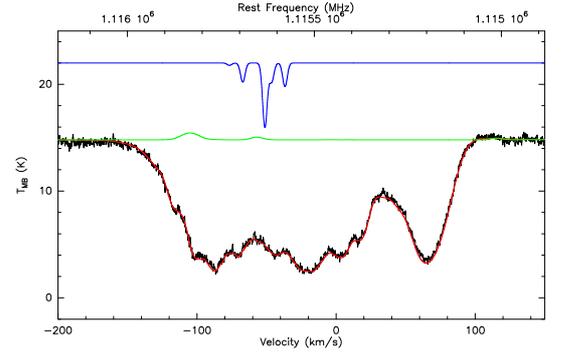}
   \caption{\htwopo{}, as already shown in \citet{Ossenkopf2010}.  The data are shown in black, the fit in red, in blue the hfs pattern is depicted, and in green the predicted contamination by other molecules.}
         \label{fig1}
\end{figure}

Determining the opacities and thus column densities of absorption lines is traditionally done using the line-to-continuum ratio.  In the present case, this is not straightforward, because the \htwopo-line has hyperfine structure with closeby components (Fig.~\ref{fig1}), which distorts the simple correspondence of line-to-continuum ratio with column density at a given velocity, and also because the line background of the Sgr~B2(M) core cannot necessarily be neglected.  We therefore fitted the lines using the XCLASS\footnote{We made use of the
myXCLASS program (https://www.astro.uni-koeln.de/projects/schilke/XCLASS), which
accesses the CDMS (\citealp{CDMS_1,CDMS_2}
http://www.cdms.de) and JPL \citealp{JPL-catalog}
http://spec.jpl.nasa.gov) molecular data bases.} program, which performs an LTE fit using the molecular data discussed in Sect.~\ref{sec:spec}, using the automated fitting routine provided by MAGIX\footnote{https://www.astro.uni-koeln.de/projects/schilke/MAGIX}.  For all velocity components, an excitation temperature of 2.7~K was assumed. \new{For molecules that react strongly with H$_2$ \citep[see the discussion in][]{Black1998, Staeuber2009}, the collisional processes in diffuse gas are unimportant relative to radiative excitation in controlling the excitation temperatures of observed transitions of species with high dipole moments such as \htwop, since inelastic collisions with H, H$_2$ and electrons compete with reactive collisons.  The excitation temperature employed here may still not be entirely correct, since particularly at 1115 GHz the general FIR background of the Galaxy contributes to a radiation temperature of 4.8~K even in the vicinity of the Sun, and for the spiral arm clouds one expects similar or slightly higher values \citep{Wright1991, Paladini2007}.  Our analysis is not affected however if the excitation temperature is dominated by this radiation field, and stays significantly smaller compared with the upper level energies for the \textit{para-} and \htwopo{} ground state lines $h\nu/k$ = 29 and 54 K, respectively, which is a well justified assumption.}
The maximum opacities of the \htwopo{} line are about 2, so the lines are only moderately opaque. 

\begin{figure}
   \centering
   \includegraphics[bb=-50 100 600 600, width=0.75\columnwidth, angle=-90]{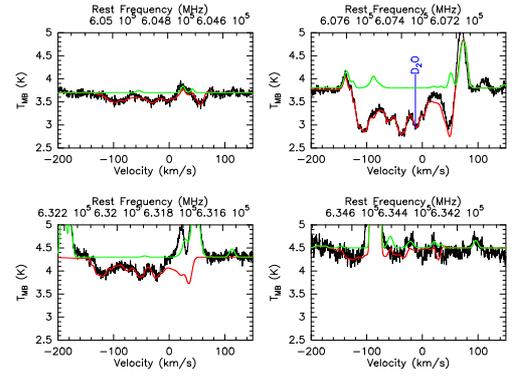}
   \caption{\htwopp{} lines in Sgr~B2, with the predicted contamination by other molecules in green, as in Fig.~\ref{fig1}.  The position of the D$_2$O ground state line is indicated.}
         \label{fig2}
\end{figure}

For \htwopp{}, only the 607~GHz line was used to perform the fit, since this is the strongest and least contaminated para-line, but it can be seen from Fig.~\ref{fig2} that predictions from this \new{reproduce} the other para lines rather well.  To predict contamination, we used the fit of all species in Sgr~B2(M) (Qin et al, in prep) as background.  This is a preliminary version of the fit, and we cannot exclude the existence of additional contamination by unknown lines.  Thus we estimate the error of the fit due to uncertainties of this nature very conservatively to be 20\%, but in the presence of strong unknown lines it could be larger at certain frequencies.  This is particularly true for the possible D$_2$O contamination of the \htwopp{}-line at 607 GHz. The relatively small variation of the \textit{ortho}/\textit{para} ratio in the absorption cloud range (see below) argues against contamination by a strong unknown line however. All \htwopp{} components are optically thin. \new{In the following, we make the assumption that the excitation of all upper levels can be approximated as LTE with an excitation temperature of 2.7~K, and that thus the \textrm{ortho-} and \htwopp{} column densities can be measured by observations of the ground state.  This assumption is reasonable for the spiral arm clouds, but most likely violated for the clouds associated with the Sgr~B2(M) envelope (see below).}

\begin{figure}
   \centering

   \includegraphics[width=0.9\columnwidth]{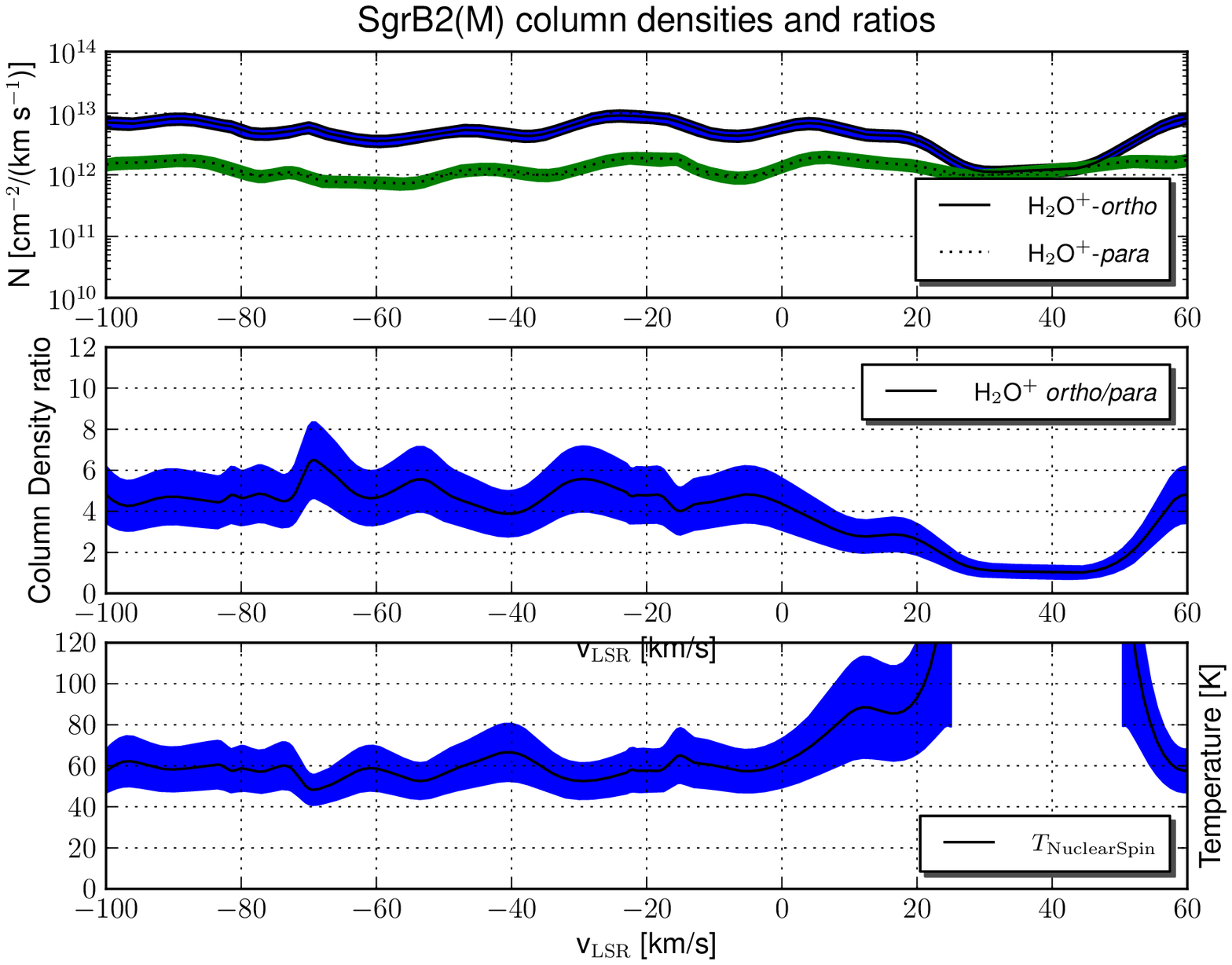}
   \caption{Column density distribution of \htwopo\ and \htwopp\ (upper panel), o/p ratio (central panel) and $T_{\rm nuclear spin}$ distribution (lower panel).}
         \label{fig3}
\end{figure}

It appears that the absorption lines of different species toward Sgr~B2(M) cannot be fitted with a unique set of physical components of fixed velocity and velocity width.  This probably reflects the different origins of the species in atomic, low density molecular and high density molecular gas with a different velocity structure. A detailed study of the different distributions will have to await the complete data set of the survey.  Apart from that, particularly in species which have hyperfine structure and are very abundant, that is in species which absorb at all velocities, the decomposition into basically Gaussian components would not necessarily be unique.  The fit rather represents a deconvolution of the hyperfine pattern.  We therefore prefer to present the results as depicted in Fig.~\ref{fig3}, as column densities/velocity interval and ratio as a function of velocity, as a sum over the components. The component of the Sgr~B2(M) envelope, which is located at 64 km s$^{-1}$, is most uncertain, because here the \htwopp{} 607 GHz line is most contaminated, and there the assumption of \new{uniformly low} excitation temperature  \new{for all levels} is most likely to be violated, since this is warm and dense gas \new{with a strong FIR field which may dominate the excitation}.

Since \htwop{} is expected to originate in mostly atomic gas at the edge of diffuse and translucent clouds, giving abundances relative to H$_2$ or H does not make sense, since it exists neither in purely atomic, nor in purely molecular gas.  \citet{Menten2010} and \citet{Qin2010} give column densities of H$_2$ of typically a few times $10^{21}$ cm$^{-2}$ and H column densities of typically 10$^{20}$ cm$^{-2}$, so the average \htwop{} abundance relative to the number of H nuclei is a few times $10^{-8}$, but could be much higher locally.   The o/p ratio was calculated using the column densities, and has a mean of 4.8 $\pm$ 0.5 for the spiral arm clouds, with little variation.  

We calculated the nuclear spin temperature using
\begin{equation}
 \frac{N(\htwopo)}{N(\htwopp)} = e^{\Delta E/kT_{\rm nuclear spin}} \frac{Q_o}{Q_p}
\end{equation}
with $\Delta E = 30.1$~K, and $Q_{o,p}$ the partition function of \htwopo/\htwopp{}, repectively, given in the Appendix.  The $Q_{o,p}$ include the rotational and nuclear spin part, and are referenced to a ground state energy of zero for both.  Toward the spiral arm clouds, we find the mean nuclear spin temperature to be almost constant at 21 $\pm 2$~K.  At this temperature range, the \new{ratio of the } partition functions \new{is} almost equal to one, and the temperature is mostly determined by the exponential factor. 

In the velocity range from about 0 to 60 km s$^{-1}$, the o/p ratio drops to unity, much below the high-temperature limit of 3. This does not reflect any thermal equilibrium and cannot be explained by any known formation mechanism: at low temperatures, the o/p ratio is extremely high, since all the molecules are in their lowest (\textit{ortho}) state, and at high temperatures the limit 3 is reached, given by the nuclear spin statistics.  We can only speculate about the cause of this unexpected result: it could be either a measurement error, or a real effect.  If it is a measurement error,  either the \htwopo{} column density is underestimated, which could be caused by contamination from a strong emission line, or the \htwopp{} column density is overestimated, which could be due to contamination from another absorption line, or the excitation temperatures deviate from the 2.7~K we assumed in such a way to produce this effect. \new{The latter could e.g.\ be produced by a bright FIR field at the location of the clouds.}  Taking the measured ratio at face value \new{as the o/p ratio} would imply that a process exists which produces \htwopo\ and \htwopp\ in equal amounts.

\section{Discussion}
Since there are no fast radiative transitions between \htwopp{} and \htwopo, the derived nuclear spin temperature is thought to be determined by chemical processes, either at formation or afterwards, since \textit{para}-\textit{ortho} transformation can only occur accompanied with a proton exchange reaction of one of the hydrogen nuclei. The only observed kinetic temperature estimates in these clouds are from \citet{Tieftrunk1994}, based on NH$_3$, and show values of 35$\pm$5~K in the --100 km s$^{-1}$ component (which is believed to originate in the Galactic center), below 20~K in the velocity range in the spiral arm clouds,  while the --10 to 20~km s$^{-1}$ component (also from the Galactic center) has temperatures exceeding 100~K \citep{Gardner1988}, and is \new{most likely} shock heated.  There is no similarity to the more or less constant nuclear spin temperature of 21~K we derive for \htwop{} formation in this range, which suggests that \htwop{} and NH$_3$ trace very different gas components: \htwop{} the warm outer photon dominated edge of clouds, NH$_3$ either the shielded and cold interior, or hot shocked gas, which also seems to be devoid of \htwop. This picture of \htwop{} formation at the edges of clouds, actually in regions where the gas is mostly atomic, is supported by \citet{Gerin2010} and \citet{Neufeld2010} based on studies of OH$+$ and \htwop{} with \textit{Herschel}/HIFI. PDR models of diffuse clouds \citep{LePetit2006}  predict temperatures of 50 to 100~K for the formation region of \htwop{} in diffuse clouds (A$_V \approx$ 1-3) with densities of about 10$^2$ and 10$^3$ cm$^{-3}$ with radiation fields of 1-3 times ambient.

The relationship of nuclear spin temperature and formation or ambient temperature needs to be discussed in somewhat more detail \citep[see][for a discussion on this ratio for molecular hydrogen]{Flower2006}.  H$_2$O$^+$ is a very reactive ion, and particularly it reacts exothermically with H$_2$ to form H$_3$O$^+$, so the reaction \htwopp\ + H$_2 \rightarrow \htwopo +$ H$_2$, which would equilibrate the o/p ratio to the kinetic gas temperature, might not be relevant. If an equivalent reaction with atomic hydrogen (\htwopp\ + H $\rightarrow \htwopo +$ H) could happen, is unknown.  If it does, it most likely will equilibrate to the current gas temperature, if not, the observed o/p ratio is the one established at formation.
How this depends on the kinetic temperature at formation is not well understood.  \htwop{} is produced by the highly exothermic reaction of OH$^+$ with H$_2$, so the variables determining the \htwop{} o/p ratio are 
\begin{enumerate}
 \item the o/p ratio of H$_2$,
\item how much and in which way the excess energy of the exothermic reaction is available for o/p conversion of \htwop,
\item the temperature of the gas at the time of formation, typically above 50-100 K based on PDR models.
\end{enumerate}
The latter two processes would push the o/p ratio down toward 3:1, the high temperature value, which means toward a spin temperature higher than the 21~K we measure.  \new{The influence of the o/p ratio of H$_2$ is hard to assess: if the reaction proceeds by way of a collision complex, then the nuclear spin of H$_2$ will have an effect on the nuclear spin of product \htwop{}, but not if the reaction proceeds through direct atom transfer.} The measured o/p of H$_2$ in diffuse clouds is about unity \citep{Savage1977, Rachford2009}. Clearly, this is an interesting area of molecular physics that needs further study. From our observations, it seems that we see an excess of \htwopo{} relative to what one would expect based on the formation temperature and available reaction energy in the spiral arm clouds and, if the measurement of an \htwopo/\htwopp\ ratio in the 0 to 60 km s$^{-1}$ region is real, an excess of \htwopp\ there.  This velocity range nominally is assigned to the Sagittarius/Scutum arms \citep{Vallee2008}, outside of the Galactic center, where no exotic conditions are expected.  However, this velocity range is also bracketed by gas local to Sgr~B2 at 0 to 10 km s$^{-1}$ and around 60 km s$^{-1}$, so it could represent diffuse gas belonging to this complex, which could be the cause of unusual excitation or chemical conditions, e.g.\ due to shocks.  In the lower velocity range overlapping with this (--10 to 20 km s$^{-1}$), \citet{Lis2010} also find water with an o/p ratio of 3, indicating high temperatures.  

\citet{Lis2010} find an average spin temperature of 27~K for water toward the the spiral arms lines of sight. H$_2$O in these clouds is formed in the gas phase, through \new{dissociative recombination} of H$_3$O$^+$, in a region where the gas is at about this temperature.  The correspondence between physical temperature and spin temperature may be more easily traced by the more stable water molecule, although in general the contribution of grain surface chemistry for water complicates the issue (see discussion in \citealt{Lis2010}).  From this study it is clear that by determining the \textit{ortho}/\textit{para}-ratio of \htwop{} (and, by proxy, from other simple hydrides) one can learn a lot about the formation processes, but also that many fundamental physical and chemical processes are still not fully understood. We can look forward to the wealth of data HIFI will bring!


\begin{acknowledgements}
  HIFI has been designed and built by a consortium of institutes and university departments from across 
Europe, Canada and the United States under the leadership of SRON Netherlands Institute for Space
Research, Groningen, The Netherlands and with major contributions from Germany, France and the US. 
Consortium members are: Canada: CSA, U.Waterloo; France: CESR, LAB, LERMA,  IRAM; Germany: 
KOSMA, MPIfR, MPS; Ireland, NUI Maynooth; Italy: ASI, IFSI-INAF, Osservatorio Astrofisico di Arcetri- 
INAF; Netherlands: SRON, TUD; Poland: CAMK, CBK; Spain: Observatorio Astron�mico Nacional (IGN), 
Centro de Astrobiolog�a (CSIC-INTA). Sweden:  Chalmers University of Technology - MC2, RSS \& GARD; 
Onsala Space Observatory; Swedish National Space Board, Stockholm University - Stockholm Observatory; 
Switzerland: ETH Zurich, FHNW; USA: Caltech, JPL, NHSC.
Support for this work was provided by NASA through an award issued by JPL/Caltech.
CSO is supported by the NSF, award AST-0540882.

H.S.P.M. is very grateful to the Bundesministerium f\"ur Bildung und 
Forschung (BMBF) for financial support aimed at maintaining the 
Cologne Database for Molecular Spectroscopy, CDMS. This support has been 
administered by the Deutsches Zentrum f\"ur Luft- und Raumfahrt (DLR). 
We appreciate funding for the ASTRONET Project CATS through the Bundesministerium f\"ur Bildung und 
Forschung (BMBF). 

We thank an anonymous referee for constructive comments which helped to clarify the discussion in this article.
\end{acknowledgements}

\Online
\begin{appendix}
\section{The spectroscopy of H$_2$O$^+$}
\label{appendix}

The rotational spectrum of oxidaniumyl was measured by laser 
magnetic resonance (LMR) \citep{H2O+_LMR_1986,H2O+_LMR_1998}; 
further infrared and electronic spectral measurements 
have been summarized in \citet{H2O+_nu2_2008}. 
Observations of the $N_{K_aK_c} = 1_{11} - 0_{00}$, 
$J = 1.5 - 0.5$ fine structure component near 1115 GHz 
with \textit{Herschel}/HIFI \citep{Ossenkopf2010} as well as subsequent 
observations raised the issue which of the two sets of 
spectroscopic parameters from LMR measurements provide 
more reliable frequency predictions. Latest  observations 
as well as reinterpretations of older ones favor 
the parameters from \citet{H2O+_LMR_1998} 
even though there seem to be small discrepancies of 
order of $\pm$5~MHz or 1.35~km s$^{-1}$ between various observations 
for the specific transition mentioned above. 
Observations carried out toward Sgr~B2(M) or Orion~KL for 
the HEXOS program are probably less suited to derive 
rest frequencies than certain other observations. 
Therefore, a preliminary catalog entry has been constructed 
for the CDMS catalog \citep{CDMS_1,CDMS_2}; the final entry 
is intended to be a common CDMS and JPL \citep{JPL-catalog} 
catalog entry.

\begin{table}
\begin{center}
\caption{Quantum numbers of rotational transitions of H$_2$O$^+$ 
described in the present work, calculated frequencies (MHz) with  
uncertainties in parentheses$^a$; lower state energies $E_{\rm lo}$ 
(K) and Einstein $A$-values (10$^{-3}$\,s${^-1}$)}
\label{lab-data}
\begin{tabular}[t]{ccr@{}lrr}
\hline \hline
\multicolumn{4}{l}{$N'_{K_a'K_c'} - N''_{K_a''K_c''}$}   & & \\[2pt]
\hline
$J' - J''$ & $F' - F''$ & \multicolumn{2}{c}{frequency} & 
\multicolumn{1}{c}{$E_{\rm lo}$} & \multicolumn{1}{c}{$A$} \\
\hline
\multicolumn{4}{l}{$1_{10} - 1_{01}$, {\it para}}  &  & \\
\hline
$1.5 - 0.5$ &     $b$     &  604678&.6~(25) & 0.005 &  1.3 \\
$1.5 - 1.5$ &     $b$     &  607227&.3~(19) & 0.000 &  6.2 \\
$0.5 - 0.5$ &     $b$     &  631724&.1~(37) & 0.005 &  5.6 \\
$0.5 - 1.5$ &     $b$     &  634272&.9~(24) & 0.000 &  2.8 \\
\hline
\multicolumn{4}{l}{$1_{11} - 0_{00}$, {\it ortho}}\\
\hline
$1.5 - 0.5$ & $1.5 - 0.5$ & 1115150&.75~(85) & 0.122 & 17.1 \\
$1.5 - 0.5$ & $0.5 - 0.5$ & 1115186&.18~(81) & 0.122 & 27.5 \\
$1.5 - 0.5$ & $2.5 - 1.5$ & 1115204&.15~(82) & 0.000 & 31.0 \\
$1.5 - 0.5$ & $1.5 - 1.5$ & 1115262&.90~(82) & 0.000 & 13.9 \\
$1.5 - 0.5$ & $0.5 - 1.5$ & 1115298&.33~(87) & 0.000 &  3.5 \\
$0.5 - 0.5$ & $0.5 - 0.5$ & 1139541&.54~(103)& 0.122 &  3.7 \\
$0.5 - 0.5$ & $1.5 - 0.5$ & 1139560&.58~(94) & 0.122 & 14.8 \\
$0.5 - 0.5$ & $0.5 - 1.5$ & 1139653&.69~(94) & 0.000 & 29.4 \\
$0.5 - 0.5$ & $1.5 - 1.5$ & 1139672&.73~(103)& 0.000 & 18.3 \\
\hline
\end{tabular}\\[2pt]
\end{center}
$^a$ Numbers in parentheses are 1\,$\sigma$ uncertainties in units of the 
least significant figures. These values should be viewed with some caution, 
in particular for the {\it para} transition, see text.\\
$^b$ $F$ is redundant for {\it para} transitions; $F = J$ may be assumed.
The lowest {\it para} state is 30.01~K above the lowest {\it ortho} state.\\
\end{table}



All infrared data and all ground state combination differences 
(GSCDs) derived from electronic spectra as summarized in 
\citet{H2O+_nu2_2008} were used in the fit as long as 
they were deemed reliable. These data are uncertain to 
between 0.005 and 0.030~cm$^{-1}$ or between 150 and 900~MHz. 
\citet{H2O+_LMR_1998} provided for their measured data 
extrapolated zero-field frequencies as well as residuals 
between observed and calculated frequencies along with uncertainties. 
From these data weighted averages of the hypothetical experimental 
zero-field frequencies and of their uncertainties were derived; 
these uncertainties were of order of 2~MHz with a considerable scatter. 
\citet{H2O+_LMR_1986} do not give sufficient data for this purpose. 
However, \citet{H2O+_LMR_1998} published calculated frequencies for 
the $1_{11} - 0_{00}$ transition. Because of the importance of 
this transition for the astronomical observation as well as 
for the fit, these calculated frequencies were also used in the fit 
with presumable uncertainties corresponding to 2~MHz. 
The quantum numbers, calculated frequencies, and uncertainties 
for the two observed rotational transitions are given in 
Table~\ref{lab-data}. Even though the LMR data fit well within 
their uncertainties, the calculated frequencies should be viewed 
with some caution because it is not clear how reliable the zero-field 
extrapolation is. Moreover, the large centrifugal distortion effects 
affecting the spectra of this ion require additional caution 
with respect to any extrapolation. It is worthwhile mentioning that 
the $1_{11} - 0_{00}$ transition frequencies in that table are 
essentially identical to the ones given in \citet{H2O+_LMR_1998}. 
The situation is different for the $1_{10} - 1_{01}$ transition. 
The $J = 1.5 - 1.5$ fine structure splitting derived from term 
values given in Table~V of \citet{H2O+_LMR_1998} differs from the 
value in Table~\ref{lab-data} by less than 4~MHz whereas the 
remaining fine structure intervals differ by up to 80~MHz. 
On the other hand, calculations directly from the \citet{H2O+_LMR_1998} 
parameters differ from values in Table~\ref{lab-data} by less than 10~MHz.
\begin{figure}[t]
\begin{center}
\includegraphics[angle=-90,width=7.5cm]{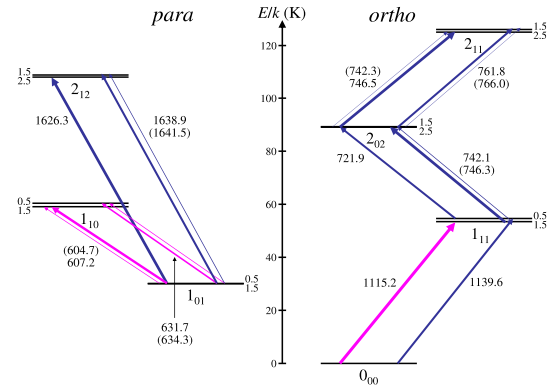}
\caption{\label{levels} Detail of the energy level diagram of H$_2$O$^+$. 
Hyperfine splitting has been omitted for the {\it ortho}-levels. 
Rotational level assignments $N_{K_aK_c}$ are given below the levels, 
fine structure level assignments $J$ to the side. Magenta arrows mark transitions 
observed in the course of the present investigation.
All other transitions shown can be observed with HIFI. The thickness 
of the arrows indicates the relative strengths of the transitions 
and the numbers the approximate frequencies. 
Frequencies of the weaker components are given in parentheses. 
The only transitions connecting to levels with $N = 1$ which are not shown 
are $2_{21} - 1_{10}$ and $2_{20} - 1_{11}$ near 2.85 and 3.0~THz 
which are not observable with HIFI but with PACS.}
\end{center}
\end{figure}
As can be seen in Fig.~\ref{levels}, the ground state {\it para} 
level $1_{01}$ of H$_2$O$^+$ is $\sim$30.13 and 30.01~K above 
the $0_{00}$ level for the $J = 0.5$ and 1.5 fine structure components 
respectively. Both {\it para} and {\it ortho} ground state levels are split 
into 2 because of the fine and hyper fine structure splitting, respectively. 
The quantum numbers are 0.5 and 1.5 in both cases, giving $g = 2$ and 4, 
respectively, and hence $Q \approx 6$ at low temperatures if {\it ortho} 
and {\it para} states are treated independently. This $1 : 1$ ratio for 
$Q$ at low temperatures approaches $3 : 1$ at room temperarture; it is about 
$1.5 : 1$ and $2 : 1$ at $\sim$40 and 75~K, respectively. 
Table~\ref{Q-p-o} gives selected partition function values for 
{\it para}- and {\it ortho}-H$_2$O$^+$, assuming they are completely 
non-interacting species. 

The mixing of {\it ortho} and {\it para} states can be mediated by terms 
such the off-diagonal electron spin-hydrogen nuclear spin coupling term $T_{ab}$ 
or the off0diagonal hydrogen nuclear spin coupling term $C_{ab} + C_{ba}$. 
The radical NH$_2$ and PH$_2$ are isoelectronic and \new{isovalent} to 
H$_2$O$^+$. Model calculations have shown the largest perturbations to 
occur between the $1_{01}$ and $1_{11}$ levels, but they are with less 
than 5 \citep{NH2_rot_2001} and less than 3~kHz \citep{PH2_rot_2002}, 
respectively, rather small, maybe even negligible. Model calculations 
suggest that perturbations of the two rotational transitions described in 
the present study are less than 3~kHz. Slightly larger perturbations 
may occur at higher quantum numbers.


\begin{table}
\begin{center}
\caption{}
\label{Q-p-o}
\begin{tabular}[t]{r@{}lr@{}lr@{}l}
\hline \hline
\multicolumn{2}{c}{$T$} & \multicolumn{2}{c}{$Q({\it para})$} & \multicolumn{2}{c}{$Q({\it ortho})$} \\
\hline
 300&.0   & 109&.1787 & 296.3567 \\
 225&.0   &  73&.4887 & 192.9372 \\
 150&.0   &  43&.0579 & 105.7512 \\
  75&.0   &  19&.0770 &  38.3647 \\
  37&.5   &  10&.4751 &  14.5910 \\
  18&.750 &   7&.3879 &   7.3124 \\
   9&.375 &   6&.2329 &   6.0584 \\
   5&.000 &   5&.9680 &   5.9982 \\
   2&.725 &   5&.9123 &   5.9961 \\
\hline
\end{tabular}\\[2pt]
\end{center}
\end{table}
\end{appendix}
\end{document}